\documentclass[aps,prl,twocolumn]{revtex4}

\begin{document}

\noindent
{\bf
Realization and Detection of Fulde-Ferrell-Larkin-Ovchinnikov Superfluid
Phases in Trapped Atomic Fermion Systems
}

In a very interesting recent Letter\cite{machida},
the authors suggested the possibility of realizing the
Fulde-Ferrell-Larkin-Ovchinnikov (FFLO)
superfluid state in trapped atomic fermion systems. This is very exciting
because the FFLO state has been of interest to condensed matter
physicists for decades, and also caught the attention of the
particle physics community recently\cite{review}. As the authors point out,
the trapped atom systems offer some unique advantages in the realization and
observation of this fascinating state. In Ref. \onlinecite{machida} the
authors used a 1D mean field solution as guidance to estimate the parameter
range for the existence of the FFLO phase, and also discussed the possibility
of its detection by imaging the atomic density of the system. In this
comment I wish to make two points. (i) In 1D there exists an exact solution
based on bosonization, which fully takes into account the important
quantum fluctuation effects\cite{yang01}; the exact solution suggests a wider
parameter range for the FFLO state than that of the mean-field
solution used in Ref. \onlinecite{machida}.
(ii) One can detect the FFLO pairing (in which Cooper pairs carry finite
momenta) more directly by extending the methods used to detect BCS
pairing\cite{regal,altman,greiner}.

(i) Based on a mean-field solution in 1D, the authors find the
following condition for the FFLO state to be stable: $\delta
n=|n_\uparrow-n_\downarrow|/n \ge \Delta_0/\pi\epsilon_F$, where
$n_\uparrow$ and $n_\downarrow$ are the densities for the two
fermion components. From this they estimate $\delta n$ needs to be
in the $10\% - 20\%$ range for the trapped atoms. While a mean-field
treatment may be justifiable for realistic quasi-1D situations, in a
genuine 1D situation (which may also be realizable experimentally)
where fluctation effects not included in mean-field theories are
severe, there exists an exact solution via
bosonization\cite{yang01}. In 1D the analog of the BCS state is the
spin gapped phase of the Luttinger liquid, in which the spin sector
is described by the gapped sine-Gorden model, and the FFLO state is
the phase with a finite soliton density in the sine-Gorden model;
the transition between the two is described by the
commensurate-incommensurate transition (CIT)\cite{yang01}. The key
point here is that the CIT is a continuous transition in which
$\delta n$ rises {\em continuously} from zero; in fact in the exact
solution $\delta n\propto \sqrt{B-B_c}$, where $B_c$ is the critical
Zeeman splitting. This suggests that in the FFLO state $\delta n$
extends all the way to zero; or when $\delta n$ itself is the
experimental control parameter, any non-zero $\delta n$ would put
the system in the FFLO phase in 1D, provided it supports a spin gap
when $\delta n=0$. This represents a significantly wider parameter
range for the FFLO state than that suggested in Ref.
\onlinecite{machida}. While only power-law long-range order is
possible in 1D, a weak 3D coupling stabilizes true long range order
for both the BCS and FFLO states, with a spatially oscillating order
parameter whose wave vector $q\propto \delta n$ for the
latter\cite{yang01}. The mean-field theory gives better descriptions
in 3D, which Ref. \onlinecite{machida} also studies.

(ii) The authors proposed to detect the FFLO state by measuring the
modulation in local magnetization, which reflects the underlying
structure of the pairing order parameter. Here we propose that we
can probe the FFLO state more directly by detecting the Cooper pairs
themselves, using the methods advanced in Refs.
\onlinecite{regal,altman,greiner}. In Ref. \onlinecite{regal} one
projects the Cooper pairs of a BCS state onto molecules by sweeping
the tuning field through the Feshbach resonance, and then use
time-of-flight (TOF) measurement to determine the molecular velocity
distribution and the condensate fraction. One can do exactly the
same experiment on the FFLO state; the fundamental difference here
is that in this case because the Cooper pairs carry intrinsic
(non-zero) momenta, the condensate will show up as peaks
corresponding to a set of {\em finite} velocities in the
distribution. An alternative method to detect the Cooper pairs is to
study the correlation in the shot noise of the fermion absorption
images in TOF\cite{greiner}, first proposed in Ref.
\onlinecite{altman}. In Ref. \onlinecite{greiner} the shot noise
correlation clearly demonstrates correlation in the occupation of
${\bf k}$ and $-{\bf k}$ states in momentum space when weakly bound
diatom molecules are dissociated. In principle the same measurement
can be performed on fermionic superfluid states, and for an FFLO
state, it would reveal correlation in the occupation of ${\bf k}$
and $-{\bf k}+{\bf q}$ states, where ${\bf q}$ is one of the momenta
of the pairing order parameter\cite{note}. Both methods allow one to
directly measure ${\bf q}$, which defines the FFLO state. They are
unique to the cold atom systems; in superconductors the only
comparable method is Josephson effect\cite{ya}.

K.Y. was supported by NSF DMR-0225698.

\vspace{12pt}
\noindent Kun Yang\\
Physics Department,
Florida State University\\
Tallahassee, Florida 32306


\end{document}